\documentclass[twocolumn, tighten]{aastex62}

\graphicspath{{./}{figures/}}

\received{June, 2020}
\revised{July, 2020}
\accepted{July, 2020}
\submitjournal{ApJ}

\shorttitle{TriAnd overdensity}
\shortauthors{Sales Silva et al.}

\begin{document}

\title{The Abundance Pattern of $\alpha$ elements in the Triangulum-Andromeda Overdensity}

\correspondingauthor{J.V. Sales Silva}
\email{joaovictor@on.br, joaovsaless@gmail.com}

\author{J. V. Sales Silva}
\affil{Observat\'orio Nacional/MCTIC, R. Gen. Jos\'e Cristino, 77,  20921-400, Rio de Janeiro, Brazil}

\author{K. Cunha}
\affil{Observat\'orio Nacional/MCTIC, R. Gen. Jos\'e Cristino, 77,  20921-400, Rio de Janeiro, Brazil}
\affiliation{Steward Observatory, University of Arizona Tucson AZ 85719} 

\author[0000-0002-0537-4146]{H. D. Perottoni}
\affiliation{Universidade Federal do Rio de Janeiro, Observat\'orio do Valongo, Lad. Pedro Ant\^onio 43, 20080-090, Rio de Janeiro, Brazil}
\affiliation{Departamento de Astronomia, IAG, Universidade de S\~ao Paulo, Rua do Mat\~ao, 1226, 05509-900, S\~ao Paulo, Brazil}

\author[0000-0002-5274-4955]{H. J. Rocha-Pinto}
\affiliation{Universidade Federal do Rio de Janeiro, Observat\'orio do Valongo, Lad. Pedro Ant\^onio 43, 20080-090, Rio de Janeiro, Brazil}

\author{S. Daflon}
\affil{Observat\'orio Nacional/MCTIC, R. Gen. Jos\'e Cristino, 77,  20921-400, Rio de Janeiro, Brazil}

\author{F. Almeida-Fernandes}
\affiliation{Universidade Federal do Rio de Janeiro, Observat\'orio do Valongo, Lad. Pedro Ant\^onio 43, 20080-090, Rio de Janeiro, Brazil}
\affiliation{Departamento de Astronomia, IAG, Universidade de S\~ao Paulo, Rua do Mat\~ao, 1226, 05509-900, S\~ao Paulo, Brazil}

\author[0000-0002-7883-5425]{Diogo Souto}
\affil{Observat\'orio Nacional/MCTIC, R. Gen. Jos\'e Cristino, 77,  20921-400, Rio de Janeiro, Brazil}
\affiliation{Departamento de F\'isica, Universidade Federal de Sergipe, Av. Marechal Rondon, S/N, 49000-000 S\~ao Crist\'ov\~ao, SE, Brazil}

\author{S. R. Majewski}
\affil{Department of Astronomy, University of Virginia, Charlottesville, VA 22904-4325, USA}



\begin{abstract}

The close relationship between the nature of the Triangulum-Andromeda (TriAnd) overdensity and the Galactic disk has become increasingly evident in recent years. However, the chemical pattern of this overdensity (R$_{GC}$ = 20 - 30 kpc) is unique and differs from what we know of the local disk. In this study, we analyze the chemical abundances of five $\alpha$ elements (Mg, O, Si, Ca, and Ti) in a sample of stars belonging to the TriAnd overdensity, including stars with [Fe/H] $<$ $-$1.2, to investigate the evolution of the $\alpha$ elements with  metallicity. High-resolution spectra from Gemini North with GRACES were analyzed. Overall, the TriAnd population presents an $\alpha$-element pattern that differs from that of the local disk; the TriAnd stars fall in between the local disk and the dwarf galaxies in the [X/Fe] vs. [Fe/H] plane. The high [Mg/Fe] ratios obtained for the lower metallicity TriAnd stars may indicate a roughly parallel sequence to the Milky Way local disk at lower values of [Fe/H], revealing a 'knee' shifted towards lower metallicities for the TriAnd population. Similar behavior is also exhibited in the [Ca/Fe] and [Si/Fe] ratios. However, for O and Ti the behavior of the [X/Fe] ratios shows a slight decay with decreasing metallicity. Our results reinforce the TriAnd overdensity as a unique stellar population of the Milky Way, with an abundance pattern that is different from all stellar populations studied to date. The complete understanding of the complex TriAnd population will require high-resolution spectroscopic observations of a larger sample of TriAnd stars.

\end{abstract}



\section{Introduction} \label{sec:intro}

The chemical abundance pattern of $\alpha$-elements (Mg, O, Si, Ca, and Ti) can reveal important characteristics of stellar populations in Milky Way. In the Galactic disk, the $\alpha$-elements abundances relative to iron present a dichotomy, with high [$\alpha$/Fe] ratios corresponding to the chemical thick disk, and low [$\alpha$/Fe] ratios corresponding to the thin disk (overall younger disk population). The relation between the [$\alpha$/Fe] ratio and [Fe/H] in the disk also shows a "knee", where the [$\alpha$/Fe] ratio decreases as the metallicity increase towards solar value, due to the onset of significant contributions from SNe Ia to interstellar enrichment (e. g., \citealt{Matteucci2003}). Recent results from large high-resolution surveys, such as SDSS III/APOGEE and SDSS IV/APOGEE-2 (\citealt{Majewski2017}), Gaia-ESO (\citealt{Gilmore2012}) and GALAH (\citealt{DeSilva2015}), indicate that the correlation between metallicity and [$\alpha$/Fe] ratios varies throughout the Galaxy, away from the mid-plane and outwards from the Galactic center (e.g. \citealt{RecioBlanco2014}, \citealt{Hayden2015}, \citealt{Weinberg2019}, \citealt{Magrini2017}). Chemodynamical simulations of the Milky Way disk also show that the relation between the [$\alpha$/Fe] and [Fe/H] ratios are expected to change for different galactocentric distances and height above the plane (e.g. \citealt{Minchev2014}, \citealt{Sharma2020}).
In addition, the "knee" position is found to vary for different dwarf spheroidal galaxies, a reflection of the wide variety of star formation histories in such systems (\citealt{Venn2004}; \citealt{Tolstoy2009}). While the level of [$\alpha$/Fe] ratio in more metal-poor stars than the "knee" metallicity depends on the IMF and preferential loss of $\alpha$ elements by selective galactic winds, for example.

Thus, given this diagnostic power, probing the behavior of [$\alpha$/Fe] versus [Fe/H] in different stellar populations in the Milky Way is key to understanding how the Galaxy assembles and evolves. One such population that helps us understand our Galaxy is the TriAngulum-Andromeda (TriAnd) overdensity (e. g. \citealt{Deason2014}, \citealt{Sheffield2014}, \citealt{Price2015}, \citealt{Xu2015}, \citealt{Li2017}, \citealt{Sheffield2018}, \citealt{Helmi2020}). 
Situated in the second Galactic quadrant about 7 kpc below the galactic plane (\citealt{Hayes2018}), the TriAnd overdensity covers $100^{\circ} < l < 150^{\circ}$ and $-15^{\circ} > b > -35^{\circ}$ (\citealt{RP2004,Deason2014,Sheffield2014,Perottoni2018}) and was discovered in parallel by \citet{RP2004} and \citet{Majewski2004} using large photometric survey data from 2MASS, and using the Washington M, T2, DDO51 system, respectively. The complex nature of the TriAnd overdensity started to be unraveled only in more recent years by high-resolution spectroscopic studies (\citealt{Bergemann2018}, \citealt{Hayes2018}, \citealt{SalesSilva2019}), which have concluded that the origin of the TriAnd system is likely linked to the interaction of the galactic disk with the satellite galaxy. However, the results are not in complete harmony.

In the first study to use high-resolution spectroscopy for stars belonging to TriAnd overdensity, \citet{Chou2011} obtained a low [Ti/Fe] ratio for three TriAnd stars, a chemical pattern reminiscent of that found in dwarf galaxies, which was interpreted as an indication of an extragalactic origin for this overdensity. More recently, however, \citet{Bergemann2018} detected similarities of the chemical pattern of the TriAnd stars to the local Galactic disk in the elements O, Na, Mg, Ti, Ba, and Eu for a sample of stars with narrow metallicity range from $-$0.66$\leq$[Fe/H]$\leq-$0.44. 
Using APOGEE results from DR14 (\citealt{Abolfathi2018}), \citet{Hayes2018} observed that the chemical pattern (Mg, (C+N), K, Ca, Mn, and Ni) of the TriAnd stars would represent an extension of the radial disk metallicity gradient to TriAnd’s radius. 
In our first study of the TriAnd population, we investigated the kinematics and chemical abundances of the elements Na, Al, Ni, Cr, Fe, Ba, and Eu (\citealt{SalesSilva2019}) in a stellar sample covering a larger metallicity range ($-$1.34$\leq$[Fe/H]$\leq-$0.78) than previous high-resolution spectroscopy studies. \citet{SalesSilva2019} concluded that the stellar population in TriAnd presents disk-like orbits and a unique chemical pattern that does not entirely resemble the full abundance pattern observed for the stars in the local Galactic disk, nor dwarf galaxies. 

Here, we add to that picture the abundance determination of Mg, O, Si, Ca, and Ti for the same seven TriAnd stars studied in \citet{SalesSilva2019} in order to try to advance the understanding of the complex nature of TriAnd overdensity stars by probing the [$\alpha$/Fe] abundance pattern. Most relevant is the fact that our sample has two low metallicity bonafide members of TriAnd ([Fe/H]$\sim$-1.3), enabling the study of $\alpha$ elements in a metallicity range never before studied in the TriAnd stellar population. 
In addition, the targets in this study also include stars that were found not to be members of TriAnd, which serve as a control sample for comparison to TriAnd stars, as well as to help gauge Galactic disk chemical trends with radius.

This paper is organized as follows: The next section describes the methodology to obtain the chemical abundances of the studied $\alpha$ elements. In Section 3, we present the results and discuss the chemical pattern of the TriAnd population (i.e., the [X/Fe] and [$\alpha_{h/ex}$] ratios) in the context of the Galaxy and its neighbors. Concluding remarks can be found in Section 4.

\section{Targets \& Methodology} \label{sec:sample}

The targets analyzed here are the same as in our previous study (\citealt{SalesSilva2019}). We selected candidate TriAnd stars from the 2MASS catalog using the combination of a color criterion that segregates giant stars from dwarf stars in the J-H versus H-K diagram (we are looking for giant stars from TriAnd) and the TriAnd overdensities regions from \citet{Perottoni2018}. Thirteen candidate stars were observed with the Gemini telescope using the GRACES spectrograph to obtain high-resolution spectra (R = 40,000) covering the optical region ($\lambda$ $\sim$ 4,000 to 10,000 \AA). We also analyzed an additional star from \citet{Bergemann2018} using a spectrum obtained from the ESO-Archive\footnote{http://archive.eso.org/}.
Based on kinematics (orbits and proper motion), \citet{SalesSilva2019} confirmed that seven out of the fourteen observed stars are members of the TriAnd overdensity. 

The atmospheric parameters ($T_{eff}$, $\log{g}$, $\xi$, and [Fe/H]) for all target stars were computed previously in \citet{SalesSilva2019}; these are presented in Table \ref{table:parameters}. We used the MARCS atmospheric models (\citealt{Gustafsson2008}) and LTE code MOOG 
(\citealt{Sneden1973}) to obtain the atmospheric parameters and compute the chemical abundances of Fe; the same code was used here to compute abundances of the chemical elements Mg, O, Si, Ca, and Ti. 
A unique solution for the atmospheric parameters was obtained using the excitation and ionization equilibrium approximations and the equivalent width independence with abundance of Fe I and Fe II lines. We also determined the atmospheric parameters for the Sun and Arcturus to test our methodologies, finding similar values to the parameters obtained in the literature for both stars (\citealt{SalesSilva2019}). Additional details about the methodology can be found in \citet{SalesSilva2019}.

\subsection{$\alpha$-element Abundance Determination}

The abundances of the $\alpha$ elements in this study were determined using equivalent width measurements (for Mg, Si, Ca, and Ti) and spectral synthesis (for O). 
In Table \ref{tabellinesa}, we present the adopted Mg I, Si I, Ca I, and Ti I line lists, along with the atomic parameters for the transitions (laboratory $gf$-values and excitation potentials obtained from \citealt{Heiter2015} and \citealt{SalesSilva2016}) and their respective equivalent width measurements. 
For the oxygen abundance derivation, we used the O I forbidden absorption line at $\lambda$ 6300.3\AA~, having $\chi$=0.0 eV and adopted $\log{gf}$=$-$9.717 (\citealt{AllendePrieto2001}). 
Our line list for computation of spectral synthesis in the [O I] region is the same line list used in \citet{SalesSilva2014} that includes the absorption line of Ni I in 6300.336\AA~ ($\log{gf}$=$-$2.31, \citealt{AllendePrieto2001}), which provides a very small contribution to the O line.


Table \ref{table:parameters} contains the LTE $\alpha$-element abundance results obtained in this study for the fourteen target stars. 
In Table \ref{sun} we present the derived $\alpha$-element abundances for the Sun and the reference red-giant Arcturus using the same line list and same 1D-LTE methodology; we also present selected solar and Arcturus abundance results from the literature. 
It can be seen that the solar abundances obtained here for Fe, Mg, Si, Ca and Ti overall compare well with the results from
\citet{GrevesseSauval1999}, and \citet{Asplund2009}. For oxygen, however, there is a larger abundance discrepancy between the studies. In particular there is a difference between the literature results depending, among other things, on whether the analysis used 1D versus 3D modeling (\citealt{Caffau2011}). We note that our solar oxygen scale has a lower abundance (A(O)$_{Sun}$=8.56).
For Arcturus, our abundance results compare well with those from \citet{RamirezAllendePrieto2011} and \citet{Jofre2015}. 

As a comparison of abundance scales, we also analyzed one star from the \citet{Bergemann2018} sample (2MASS23174139+3113043; star 14) using the same spectrum from the ESO archive they analyzed (see details in \citealt{SalesSilva2019}). \citet{Bergemann2018} determined the NLTE abundance of Mg and Ti for the star 14. Here we find similar values for the abundance ratios of Fe, Mg and Ti in NLTE as derived in \citet{Bergemann2018}, with a difference (This study - \citealt{Bergemann2018}) of $\Delta$[Fe/H]=0.02, $\Delta$[Mg/Fe]= $-$0.09 and $\Delta$[Ti/Fe]= $-$0.11.


In order to assess possible departures from LTE for the derived abundances we estimate NLTE corrections for our star sample. Based on the calculations presented in \citet{Bergemann2013} and \citet{Bergemann2015}, the NLTE corrections for the selected transitions of Mg I, Si I and O I in this study are deemed to be negligible given the line list and the range in the atmospheric parameters of our star sample. 
For estimating NLTE abundances for Ca and Ti we use the studies by \citet{Mashonkina2007} and \citet{Bergemann2011}, respectively. 
\citet{Bergemann2011} built a Ti model atom with 216 energy levels to calculate NLTE departures for a large number Ti I absorption lines, including the fourteen Ti I lines measured here;  \citet{Mashonkina2007} used a model atom with 63 energy levels to define the NLTE corrections for a number of Ca transitions and these include the four Ca I lines studied here.  
Assuming that the NLTE corrections derived in these studies are valid to apply to our LTE abundances (which is a reasonable assumption given that they were computed for the same family of model atmospheres), we can correct our individual line abundances, line by line, and then compute the average Ti and Ca NLTE abundances for each star (Table \ref{table:parameters}). 
On average, the difference between the NLTE and LTE abundances for our sample is 0.10$\pm$0.03 dex for Ca and 0.09$\pm$0.01 dex for Ti. Table \ref{table:parameters} also contains the NLTE [X/Fe] ratios for Ca and Ti for the target stars.

\subsection{Abundance Uncertainties}

The atmospheric parameter uncertainties for the stars of our sample are 75 K, 0.2 dex and 0.1 km/s for $T_{eff}$, $\log{g}$ and $\xi$, respectively, as discussed previously in \citet{SalesSilva2019}. 
In Table \ref{error}, we present the abundance uncertainties regarding the target star 2MASS02510349+4342045, taken as typical in our sample. To calculate the abundance sensitivities we considered the abundance variations caused independently by each atmospheric parameter, varying these parameters by their respective uncertainties; we determined the final uncertainties of chemical abundances adding in quadrature the uncertainties in abundance relative to each atmospheric parameter. 

\begin{table*} 
\tabcolsep 0.17truecm
\tiny
\caption{Target Stars, Stellar Parameters \& Abundances}
\begin{tabular}{lcccccccccccc}\hline\hline
\multicolumn{13}{c}{TriAnd stars}\\\hline
& 2MASS ID &  $T_{eff}$  & $\log{g}$   & $\xi$    & [Fe\,{\sc i}/H]     &  [Mg/Fe]        &  [Si/Fe]      &   [Ca/Fe]$_{LTE}$ &   [Ca/Fe]$_{NLTE}$&  [Ti/Fe]$_{LTE}$& [Ti/Fe]$_{NLTE}$&   [O/Fe]  \\\hline
\# & &     K              &                & km/s         &                                    &                                     &        &         &       &        &     &         \\\hline
3 & 00594094+4614332 &  4100              &  0.4           &  1.77                &  $-$0.82$\pm$0.10   &  0.13$\pm$0.15  & 0.21$\pm$0.09 &  0.12$\pm$0.09  &  0.22$\pm$0.09  & 0.21$\pm$0.11 & 0.32$\pm$0.11 &   0.29    \\
5 & 01151944+4713512 &  4075              &  0.7           &  1.99                &  $-$0.94$\pm$0.10   &  0.27$\pm$0.14  & 0.19          &  0.17$\pm$0.16  &  0.27$\pm$0.16  & 0.26$\pm$0.12 & 0.34$\pm$0.12 &   0.42    \\
6 & 02485891+4312154 &  3900              &  0.5           &  1.82                &  $-$0.81$\pm$0.13   &  0.15$\pm$0.15  & 0.34$\pm$0.13 & -0.06  &  0.04      & 0.05$\pm$0.10 & 0.13$\pm$0.10 &   0.25    \\
7 & 23535441+3449575 &  4200              &  1.3           &  0.60                &  $-$0.78$\pm$0.10   & -0.06$\pm$0.07  & 0.19$\pm$0.10 &  0.12$\pm$0.09  &  0.22$\pm$0.09  & 0.06$\pm$0.14 & 0.16$\pm$0.14 &   0.38    \\
9 & 02350813+4455263 &  4050              &  0.5           &  0.59                &  $-$1.23$\pm$0.09   &  0.21$\pm$0.09  & 0.40$\pm$0.10 &  0.09$\pm$0.16  &  0.19$\pm$0.16  &-0.08$\pm$0.12 & 0.01$\pm$0.12 &   0.11    \\
11 & 02510349+4342045 &  4025              &  0.9           &  1.63                &  $-$0.78$\pm$0.10   &  0.15$\pm$0.06  & 0.26$\pm$0.08 &  0.01$\pm$0.13  &  0.11$\pm$0.13  & 0.01$\pm$0.13 & 0.11$\pm$0.13 &   0.29    \\
12 & 02475442+4429269 &  4000              &  0.3           &  0.97                &  $-$1.34$\pm$0.12   &  0.34$\pm$0.07  & 0.39$\pm$0.07 &  0.09$\pm$0.16  &  0.19$\pm$0.16  &-0.01$\pm$0.12 & 0.10$\pm$0.12 &   0.27    \\\hline
\multicolumn{13}{c}{Non-TriAnd stars}\\\hline
1 & 00075751+3359414 &  4150              &  0.6           &  0.88                &  $-$1.71$\pm$0.09   &   0.25$\pm$0.14 & 0.20$\pm$0.10 &  0.36$\pm$0.08  &  0.45$\pm$0.08  & 0.05$\pm$0.11 & 0.12$\pm$0.11 &   0.63    \\
2 & 00534976+4626089 &  3925              &  1.3           &  1.81                &  $-$0.46$\pm$0.10   &   0.05$\pm$0.03 & 0.38$\pm$0.11 & -0.33$\pm$0.09  & -0.29$\pm$0.09  &-0.07$\pm$0.12 &-0.01$\pm$0.12 &   0.25    \\
4 & 01020943+4643251  &  4125              &  0.0           &  1.04                &  $-$1.50$\pm$0.11   &   0.30$\pm$0.19 & 0.45$\pm$0.13 &  0.23$\pm$0.06  &  0.33$\pm$0.06  & 0.14$\pm$0.12 & 0.25$\pm$0.12 &   0.57    \\
8 & 23481637+3129372 &  3975              &  0.4           &  1.93                &  $-$1.42$\pm$0.08   &   0.22$\pm$0.20 & 0.59$\pm$0.08 &  0.22$\pm$0.14  &  0.29$\pm$0.14  & 0.28$\pm$0.10 & 0.37$\pm$0.10 &   0.67    \\
10 & 23495808+3445569 &  3925              &  1.4           &  1.96                &  $-$0.63$\pm$0.10   &  -0.01$\pm$0.02 & 0.35$\pm$0.13 & -0.13$\pm$0.18  & -0.02$\pm$0.18  &-0.09$\pm$0.09 &-0.02$\pm$0.09 &   0.58    \\
13 & 02463235+4314481 &  4100              &  0.6           &  1.98                &  $-$1.03$\pm$0.08   &   0.13$\pm$0.13 & 0.44$\pm$0.11 &  0.25$\pm$0.13  &  0.35$\pm$0.13  & 0.34$\pm$0.11 & 0.43$\pm$0.11 &   0.49    \\
 14 & 23174139+3113043 &  3925           &  0.3           &  1.62           &  $-$0.91$\pm$0.13   &   0.15          & ---           &  0.03$\pm$0.14  &  0.22$\pm$0.14  & 0.14$\pm$0.13 & 0.22$\pm$0.13 &    ---    \\\hline
\hline                                                                 
\end{tabular}
\label{table:parameters}
\end{table*}

\begin{table*}
\caption{Line Lists \& Measured Equivalent Widths}
\footnotesize
\begin{tabular}{ccccccccccccccccccc}
\label{tabellinesa}
\\\tableline\tableline
    & & & & &\multicolumn{14}{c}{Equivalent Widths (m\AA)} \\\tableline
\multicolumn{5}{c}{} & \multicolumn{14}{c}{Star}\\
\cline{6-19}
El. & $\lambda$ & $\chi$(eV) & $\log{gf}$ & Ref & 1 & 2 & 3 & 4 & 5 & 6 & 7 & 8 & 9 & 10 & 11 & 12 & 13 & 14\\
\tableline                                                                                              
Mg\,{\sc i} & 4730.04 & 4.34 & $-$2.390 & R03  &  23 &  96 & --- &  28 & --- & --- & --- & --- & --- & --- & --- &  85 &  64 & --- \\
Mg\,{\sc i} & 6318.71 & 5.11 & $-$1.940 & Ca07 & --- &  64 &  63 &  26 &  54 &  57 &  37 &  35 &  33 &  48 &  48 &  59 &  44 &  40 \\
Mg\,{\sc i} & 6319.24 & 5.11 & $-$2.160 & Ca07 &  10 &  47 &  35 &  10 &  35 &  33 &  23 &  14 &  19 &  35 &  35 &  36 &  35 & --- \\
Mg\,{\sc i} & 6319.49 & 5.11 & $-$2.670 & Ca07 &   4 & --- & --- & --- &  23 &  23 & --- & --- &  10 &  15 &  15 &  17 & --- & --- \\
Si\,{\sc i} & 5793.08 & 4.93 & $-$2.060 & R03  & --- &  54 &  45 &  25 &  36 &  52 &  32 &  38 &  36 &  45 &  45 &  44 &  47 & --- \\
Si\,{\sc i} & 6125.03 & 5.61 & $-$1.540 & E93  & --- & --- & --- & --- & --- & --- & --- & --- & --- & --- & --- & --- &  35 & --- \\
Si\,{\sc i} & 6131.58 & 5.62 & $-$1.680 & E93  &   5 &  28 & 26 &  --- &--- &  21 &  14 &  --- & 13 &  --- &--- &  19  & --- & --- \\ 
Si\,{\sc i} & 6145.02 & 5.61 & $-$1.430 & E93  &   6 & --- & --- &  10 & --- & --- &  23 & --- &  20 &  23 &  23 &  28 &  32 & --- \\
Si\,{\sc i} & 7760.64 & 6.20 & $-$1.280 & E93  & --- & --- & --- & --- & --- &  11 &  13 & --- & --- & --- & --- & --- & --- & --- \\
Si\,{\sc i} & 8728.01 & 6.18 & $-$0.360 & E93  &  19 &  44 &  50 &  30 &  42 &  37 &  39 &  38 &  33 &  45 &  45 &  38 &  48 & --- \\
Ca\,{\sc i} & 6161.30 & 2.52 & $-$1.270 & E93  &  58 & --- & 131 &  66 & 136 & 133 &  83 & --- &  74 & 143 & 143 & 126 & 130 & 121 \\
Ca\,{\sc i} & 6166.44 & 2.52 & $-$1.140 & R03  &  60 & 139 & 132 &  68 & 140 & 133 &  82 & 120 &  71 & --- & --- & 124 & 140 & 111 \\
Ca\,{\sc i} & 6169.04 & 2.52 & $-$0.800 & R03  &  78 & --- & --- &  85 & --- & --- & 101 & 130 &  82 & --- & --- & 145 & --- & 145 \\
Ca\,{\sc i} & 6455.60 & 2.51 & $-$1.290 & R03  &  65 & 126 & 124 &  70 & 118 & --- &  87 & --- &  78 & 129 & 129 & 110 & 118 & 117 \\
Ti\,{\sc i} & 4758.12 & 2.25 &    0.510 & L13  & --- & --- & 120 & --- & --- & --- & --- & --- & --- & --- & --- & --- & --- & --- \\
Ti\,{\sc i} & 4759.27 & 2.26 &    0.590 & L13  & --- & --- & --- & --- & --- & 132 & --- & --- & --- & --- & --- & 109 & --- & --- \\
Ti\,{\sc i} & 5043.58 & 0.84 & $-$1.590 & L13  & --- & --- & --- & --- & --- & --- & --- & 127 & --- & --- & --- & 126 & 148 & --- \\
Ti\,{\sc i} & 5062.10 & 2.16 & $-$0.390 & L13  &  23 & --- & --- & --- & --- & --- & --- &  69 &  40 & 110 & 110 & --- & 100 &  92 \\
Ti\,{\sc i} & 5113.44 & 1.44 & $-$0.700 & L13  &  49 & --- & --- &  66 & 141 & --- &  74 & --- &  64 & --- & --- & --- & --- & --- \\
Ti\,{\sc i} & 5145.46 & 1.46 & $-$0.540 & L13  &  53 & --- & 140 &  74 & --- & --- & --- & 135 &  66 & --- & --- & --- & 141 & --- \\
Ti\,{\sc i} & 5223.62 & 2.09 & $-$0.490 & N98  &  14 &  99 & --- & --- & --- & --- & --- & --- & --- & --- & --- & --- & --- &  84 \\
Ti\,{\sc i} & 5295.78 & 1.07 & $-$1.590 & L13  &  36 & 142 & 134 &  56 & 113 & --- &  60 & 126 & --- & --- & --- & 103 & 118 & --- \\
Ti\,{\sc i} & 5490.15 & 1.46 & $-$0.840 & L13  &  45 & --- & 139 &  62 & --- & --- &  71 & 126 & --- & 143 & 143 & --- & --- & 132 \\
Ti\,{\sc i} & 5662.15 & 2.32 &    0.010 & N98  & --- & 127 & 111 &  53 & 114 & --- &  64 &  96 & --- & 115 & 115 & 104 & 119 & 101 \\
Ti\,{\sc i} & 5689.46 & 2.30 & $-$0.360 & N98  & --- & 106 &  96 &  32 &  97 & 102 &  54 &  60 &  38 &  94 &  94 &  83 &  84 &  79 \\
Ti\,{\sc i} & 5922.11 & 1.05 & $-$1.380 & L13  &  59 & --- & --- & --- & --- & --- &  88 & 140 & --- & --- & --- & --- & --- & --- \\
Ti\,{\sc i} & 5978.54 & 1.87 & $-$0.440 & B86  &  47 & 138 & 128 &  67 & 128 & 131 &  76 & 107 &  67 & --- & --- & 115 & 128 & 120 \\
Ti\,{\sc i} & 6091.17 & 2.27 & $-$0.320 & L13  & --- & 112 &  89 &  40 & 109 & 115 &  63 &  73 & --- & 102 & 102 &  98 &  97 &  96 \\
Ti\,{\sc i} & 6126.22 & 1.07 & $-$1.368 & B83  & --- & --- & --- & --- & --- & --- & --- & 141 & --- & --- & --- & --- & --- & --- \\
Ti\,{\sc i} & 6554.22 & 1.44 & $-$1.150 & L13  &  43 & --- & 138 & --- & 148 & --- &  72 & 115 &  68 & --- & --- & 126 & 131 & 147 \\                    
\tableline                                                          
\end{tabular} 
\\
References for $\log{gf}$ values: B83: \citet{Blackwell1983}; B86: \citet{Blackwell1986}; Ca07: \citet{Carretta2007}; E93: \citet{Edvardsson1993};\ L13: \citet{Lawler2013}; N98: \citet{Nitz1998}; R03: \citet{Reddy2003}
                 
\end{table*}

\begin{table}
\caption{Arcturus and Solar abundances.}
\tabcolsep 0.07truecm
\tiny
\label{sun}
\begin{tabular}{lccc|ccc}\\\hline\hline
\multicolumn{4}{c}{Sun} & \multicolumn{2}{c}{Arcturus}\\\hline
El.              & This  & Grevesse \&  & Asplund        & This  & Ramirez \& Allende    &    Jofr\'e    \\
$_{\rule{0pt}{8pt}}$ & work  & Sauval (1999)& et al. (2009)  & work  &  Prieto (2011) & et al. (2015)\\
\hline     
Fe                   & 7.50  & 7.50         & 7.50           & 6.91  & 6.98   &    6.92    \\ 
O                    & 8.56  & 8.83         & 8.69           & 8.56  & 8.66    &   ---     \\
Mg                   & 7.61  & 7.58         & 7.60           & 7.45  & 7.47   &   7.49  \\
Si                   & 7.60  & 7.55         & 7.51           & 7.27  & 7.30     &   7.24  \\
Ca                   & 6.40  & 6.36         & 6.34           & 6.02  & 5.94    &   5.92  \\
Ti                   & 4.94  & 5.02         & 4.95           & 4.68  & 4.66      &   4.59  \\
\hline
\end{tabular}
\\\textbf{Notes.} A(0)$_{Sun}$=8.76 in \citet{Caffau2011}.
\end{table}


\begin{table}
\caption{Abundance Uncertainties}
\label{error}
\begin{tabular}{lcccc}\\\hline\hline
[X/H] & $\Delta T_{eff}$ & $\Delta\log{g}$ & $\Delta\xi$ & $\left( \sum \sigma^2 \right)^{1/2}$ \\
$_{\rule{0pt}{8pt}}$ & $+$75~K & $+$0.2 & $+$0.1 km\,s$^{-1}$ &  \\
\hline     
Fe\,{\sc i}    & $-$0.01  & $+$0.05 & $-$0.05 & 0.07 \\ 
O\,{\sc i}     &    0.00  & $+$0.08 & $-$0.02 & 0.08 \\
Mg\,{\sc i}    &    0.00  & $+$0.02 & $-$0.02 & 0.03 \\
Si\,{\sc i}    & $-$0.08  & $+$0.06 & $-$0.01 & 0.10 \\
Ca\,{\sc i}    & $+$0.08  & $+$0.01 & $-$0.08 & 0.11 \\
Ti\,{\sc i}    & $+$0.10  & $+$0.02 & $-$0.08 & 0.13 \\
\hline\\
\end{tabular}

\par \textbf{Notes.} The abundance sensitivities were computed for the star 2MASS02510349+4342045. Each column gives the variation of the abundance caused by the variation in $T_{eff}$, $\log{g}$ and $\xi$. The last column gives the net RMS uncertainty in the abundance after combination of the uncertainties in columns 2 through 4.

\end{table}

\section{Discussion} \label{sec:resu}




    In \citet{SalesSilva2019}, we carried out our first high-resolution spectroscopic study of stars belonging to TriAnd overdensity, deriving the abundances of Na, Al, Fe, Cr, Ni, and the heavy elements Ba and Eu. That study found that the TriAnd population had a unique chemical pattern that did not match that of the local disk, or of dwarf galaxies.
    Here, we continue our investigation of the TriAnd region by deriving the abundance patterns of five $\alpha$-elements ~\textemdash~ O, Mg, Si, Ca, and Ti ~\textemdash~ in the same stars studied previously in \citet{SalesSilva2019}. 
    
    \subsection{$\alpha$-Elements in the Triangulum-Andromeda overdensity} 
    
    The TriAnd targets analyzed are all metal-poor, encompassing a metallicity range between [Fe/H]=$-$0.78 and $-$1.34; in this study we find that most of the TriAnd targets analyzed show an overabundance of $\alpha$ elements relative to Fe (Table \ref{table:parameters}). 
    
\begin{figure*}
\centering
\includegraphics[width=15cm]{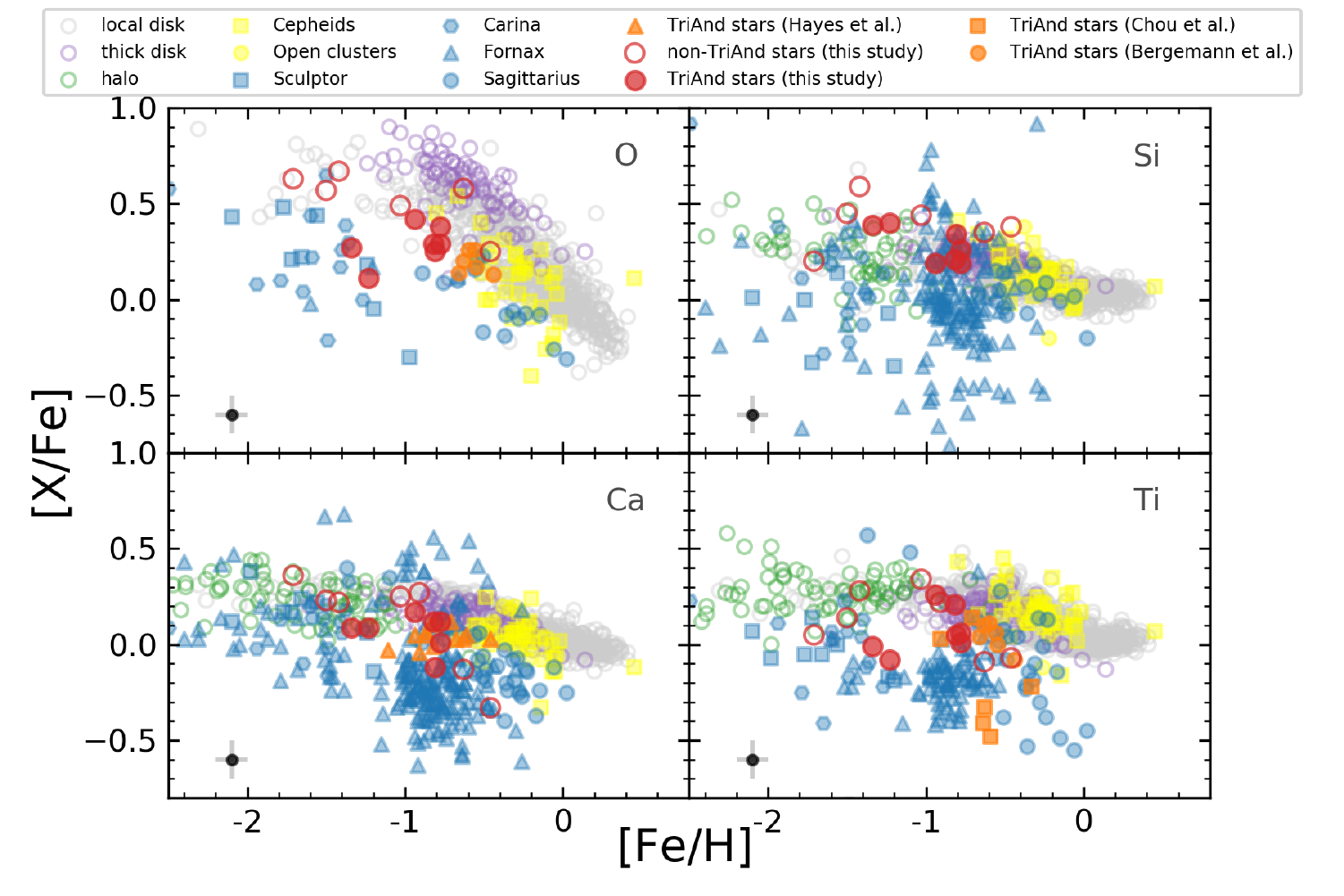}
\caption{The abundance ratios of [O, Si, Ca and Ti/Fe] vs. [Fe/H] in different populations. The filled red circles represent our TriAnd member targets, while the open red circles are our non-TriAnd control sample. The orange triangles are the TriAnd stars from \citet{Hayes2018}; the orange circles are the TriAnd stars from \citet{Bergemann2018}; the orange squares are the TriAnd stars from \citet{Chou2011}. Results for the Milky Way are also shown: the gray circles represent the local disk stars from \citet{Bensby2014}; the purple circles are the thick disk stars from \citet{Reddy2006}; the yellow circles are the open clusters from the outer disk from \citet{Yong2012}; the yellow squares are the Cepheids from outer disk from \citet{Luck2011}, \citet{Lemasle2013} and \citet{Genovali2015}; the green circles are the halo stars from \citet{Ishigaki2012}. The blue squares represent the stars from the Sculptor dwarf Galaxy from \citet{Geisler2005} and \citet{Shetrone2003};
the blue hexagons are the stars from the Carina dwarf galaxy from \citet{Koch2008} and \citet{Shetrone2003}; the blue triangles are the stars from the Fornax dwarf galaxy from \citet{Shetrone2003} and \citet{Letarte2010}; the blue circles are the stars from the Sagittarius dwarf galaxy from \citet{Monaco2005} and \citet{Sbordone2007}.
The Ca and Ti abundances shown are the ones derived here in LTE, without NLTE corrections to better compare with the LTE chemical abundances of the literature.
The error bars shown in the lower left corners of all panels represent the expected uncertainties in the derived abundance ratios. 
}
\label{singlealphaartigo4final}
\end{figure*}

\begin{figure*}
\centering
\includegraphics[width=15cm]{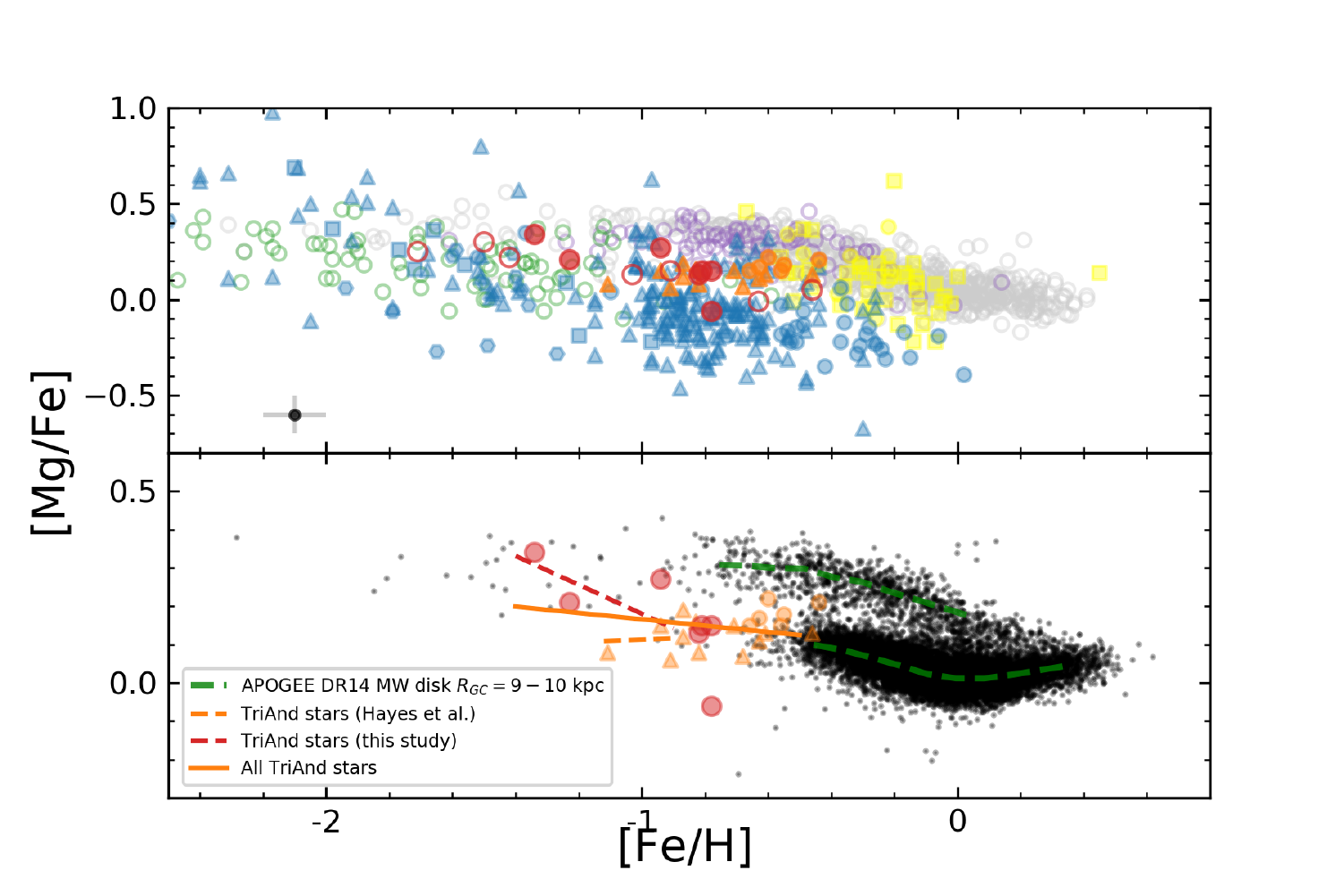}
\caption{Abundance ratios of [Mg/Fe] vs. [Fe/H]. The symbols have the same meaning as in Figure \ref{singlealphaartigo4final}. In the bottom panel of the figure we show as black dots the APOGEE DR14 results for the Milky Way in the solar neighborhood (R$_{GC}$ = 9 - 10 kpc) along with its respective median tendency (shown as the green dashed-line); the orange and red dashed-lines correspond to the trends for the TriAnd stars from \citet{Hayes2018} and our study, respectively, for [Fe/H] $< -$1.0 dex, while the solid orange line corresponds to the linear fit for all TriAnd stars (\citealt{Hayes2018}, \citealt{Bergemann2018} and our study).
}%
\label{singlealphaartigo5final}
\end{figure*}

    In the different panels of Figure \ref{singlealphaartigo4final}
  we show the [O, Si, Ca, Ti/Fe] abundance ratios versus [Fe/H] obtained in this study for the TriAnd stars (filled red circles), along with the results obtained  with the same methodology for those targets in our sample that were not found to be members of TriAnd (shown as open red circles). 
  Figure \ref{singlealphaartigo4final} also includes results for TriAnd from infrared APOGEE spectra (\citealt{Hayes2018}; orange triangles) and the two other high-resolution studies in the literature (\citealt{Chou2011}, orange squares, and \citealt{Bergemann2018}, orange circles). 
  In addition, as comparisons to other stellar populations in the Milky Way, we also show results of optical studies from the literature for the halo (\citealt{Ishigaki2012}), field stars in the local disk (\citealt{Bensby2014}) and open clusters and Cepheids in the outer disk (R$_{GC}>$ 12 kpc; \citet{Yong2012}, \citealt{Luck2011}, \citealt{Lemasle2013} and \citealt{Genovali2015}). Finally, we also show results for four dwarf galaxies: Sculptor (\citealt{Geisler2005} and \citealt{Shetrone2003}), Carina (\citealt{Koch2008} and \citealt{Shetrone2003}), Fornax (\citealt{Shetrone2003} and \citealt{Letarte2010}) and Sagittarius (\citealt{Monaco2005} and \citealt{Sbordone2007}). 
  The top panel of Figure \ref{singlealphaartigo5final} shows the corresponding plot for magnesium with the same comparison samples.

  
   The first highlight from our results in Figures \ref{singlealphaartigo4final} and \ref{singlealphaartigo5final} (top panel) is that, in general terms, the TriAnd stars have $\alpha$-element abundances (except for Si) that are overall displaced to lower values of [X/Fe] for a given [Fe/H] when compared to the results for the local Galactic trend.
   It is noticeable, however, that the derived [X/Fe] ratios for the TriAnd stars do not reach the lowest [X/Fe] ratios found in dwarf spheroidal galaxies (blue symbols); see, for example, the results for Fornax (shown as blue triangles), which overlap in metallicity range with the TriAnd stars.  
   The [Si/Fe] ratios obtained for the TriAnd stars are slightly higher when compared to the other $\alpha$ elements, being more similar to the low-metallicity end of the local disk pattern (at metallicities [Fe/H] $\sim -1$). These [Si/Fe] values also overlap those for a number of dSph stars ~\textemdash~ although the Si abundance results for the dwarf spheroidals in the figure extend to significantly lower values than those from either TriAnd or the local disk.
   
   One important aspect in this comparison is that the derived abundances for TriAnd stars are offset from the results for those targets (shown as open red symbols) that did not have a high probability to being actual TriAnd members after an analysis of their orbits and proper motions; such stars can serve as a control sample in comparison with the TriAnd stars.
   (See details of how sample stars were deemed as non-TriAnd members based on kinematical criteria in \citealt{SalesSilva2019}). 
    Figures \ref{singlealphaartigo4final} and \ref{singlealphaartigo5final} (top panel) show that the [X/Fe] ratios for the $\alpha$-elements for the non-TriAnd stars are generally higher and seem to overall follow the Milky Way halo pattern (shown as green open circles). Such distinct abundance results for the non-TriAnd stars when compared to the TriAnd stars, all coming from a homogeneous analysis, is reassuring that we may be detecting a different abundance pattern for the TriAnd population that is not simply due to systematic differences ~\textemdash~ with the caveat, of course, that this is all based on a small number of stars.
    In our previous study, \citet{SalesSilva2019} have also found that these same non-TriAnd targets were chemically segregated from the stars members of the TriAnd overdensity.

  Concerning other results for the abundances of $\alpha$ elements for TriAnd stars from the literature, Figures \ref{singlealphaartigo4final} and \ref{singlealphaartigo5final} (top panel) show that, overall, the results from \citealt{Hayes2018} and \citealt{Bergemann2018} (represented by orange triangles and circles, respectively) do not provide a dissonant picture: They generally fall below the local Milky Way pattern. 
  However, a unique contribution of our sample is that it extends our mapping of TriAnd chemistry to a more metal-poor regime.

  \subsection{Trends in the [X/Fe] vs [Fe/H] Plane}

The main sources of enrichment of $\alpha$-elements in the interstellar medium are high mass stars. Therefore, the [X/Fe] ratios and its trend for the $\alpha$-elements of a stellar population can provide important information about the relative number of high to low mass stars (i.e., the IMF) and the rate of SN II to SNe Ia in the interstellar medium that formed these stars. Also, we can understand the chemical evolution of a population probing the [X/Fe] vs [Fe/H] pattern in low metallicity stars.
In terms of iron abundances our TriAnd sample can be divided roughly into two groups, stars with [Fe/H] $> -$ 1 (the more metal-rich ones that overlap in metallicity with the TriAnd sample discussed in \citealt{Hayes2018}) and the metal-poor TriAnd stars with metallicities [Fe/H] $\sim -$1.3. 
We discuss below the apparently distinct trends observed for the differrent alpha elements studied here, which are  highly influenced by the low metallicity stars in the sample.  However, we must regard any such derived trends as tentative, given the still small number of TriAnd stars studied both here and in the literature.
    
  \subsubsection{Oxygen and Titanium}
  
   The behavior of the O and Ti abundances with metallicity seem similar (Figure \ref{singlealphaartigo4final}), and their trends with metallicity appear to be at odds with those of the Milky Way pattern.
   Starting with the O and Ti abundances for those stars with higher metallicities in our TriAnd sample, we can see that these roughly agree with the results from \citet[orange circles]{Bergemann2018}. However, the two lowest metallicity stars in our sample have lower [O, Ti/Fe] ratios, being less enhanced at low metallicity than at high metallicity.
   Such behavior does not follow the chemical pattern of the thick disk (or halo), rather, the [Ti, O/Fe] ratios for the low metallicity TriAnd stars overlap with results for dwarf galaxies (shown as blue symbols in Figure \ref{singlealphaartigo4final}). 

    \citet{Chou2011} found very low [Ti/Fe] ratios ([Ti/Fe]$<-$0.3) for three TriAnd stars in their sample (see orange squares in Figure \ref{singlealphaartigo4final}), attributing an extragalactic origin to these stars and the TriAnd overdensity. 
    None of the stars in our sample show as low [Ti/Fe] ratios (see Table \ref{table:parameters}); \citet{Bergemann2018} do not find such low [Ti/Fe] ratios either. We note that there are other stars in the \citet{Chou2011} sample that exhibit higher [Ti/Fe] ratios, more in line with what we find for our sample.

   \subsubsection{Silicon and Calcium}
 
 The Si/Fe and Ca/Fe abundances obtained for the studied TriAnd stars are enhanced relative to solar (except for one star having a negative [Ca/Fe] = $-$0.06, Table \ref{table:parameters}). 
 For silicon, we find a more 'canonical' behavior of [Si/Fe] with metallicity, or, more specifically we find that the metal-poor stars in our TriAnd sample have higher Si over Fe ratios when compared to the more metal-rich TriAnd stars.  For Ca, the behavior is similar but slightly flatter than for Si, however, not significantly distinct. The Ca abundances for the APOGEE sample analyzed in \citet{Hayes2018} (orange triangles) are in rough agreement with our results, but, as mentioned previously, their sample does not extend much below  metallicities below [Fe/H]=$-$1; there are no TriAnd results for Si in the literature.
 Therefore, considering all results the Si and Ca abundances for TriAnd seem to indicate a tendency that the most metal-poor stars ([Fe/H] $<-$1) have more enhanced [Si, Ca /Fe] ratios when compared to the more metal-rich TriAnd stars; such behavior may be interpreted as indicative that, at metallicities below [Fe/H]$\sim -$1, TriAnd stars might be before where the so-called ‘knee’ where the production of Fe from SNe Ia becomes significant, and where the [Si, Ca/Fe] abundance ratios are at a plateau, in a similar fashion to what is observed for the thick disk or halo pattern.

\subsubsection{Magnesium}
     
We find that all stars in our TriAnd sample, except for one, are enriched in Mg with respect to Fe, having a mean [Mg/Fe] of 0.17$\pm$0.13; the three lowest metallicity targets (Stars 5, 9 and 12) also have the highest [Mg/Fe] ratios (of 0.27, 0.21, and 0.34 dex, respectively; Table \ref{table:parameters}). The behavior for [Mg/Fe] versus [Fe/H] for TriAnd is similar to that of Ca and Si, whose ratios with respect to iron also increase with decreasing metallicity (see Figure \ref{singlealphaartigo5final}).

Magnesium was the only $\alpha$ element studied both in \citet{Hayes2018} and \citet{Bergemann2018}. All of the TriAnd [Mg/Fe] ratios in the latter studies (shown in Figure \ref{singlealphaartigo5final} as orange circles and triangles) are slightly enhanced relative to solar. As previously noted, the TriAnd sample in \citet{Bergemann2018} is concentrated to in a small metallicity range (-0.66 $\leq$ [Fe/H] $\leq$ -0.44) and they also find clustered magnesium abundances.

In Figure \ref{singlealphaartigo5final} (top panel) we show the [Mg/Fe] ratios as a function of [Fe/H] for TriAnd along with results for the same populations as discussed before, while in the bottom panel of Figure \ref{singlealphaartigo5final} we show as black dots the APOGEE results for the Milky Way with 9 $\leq$ R$_{GC} \leq$ 10 kpc (the same APOGEE sample of $\approx$ 14,000 stars in their upper left panel of Figure 3 shown in the \citealt{Hayes2018} study). We generally find the same pattern in the [Mg/Fe] versus [Fe/H] plane as observed for the other $\alpha$-elements: the TriAnd population is located between the Milky Way disk and the dwarf galaxy populations in the same metallicity range (see top panel in Figure \ref{singlealphaartigo5final}).

Figure \ref{singlealphaartigo5final} (bottom panel) shows that the two disk components (low and high $\alpha$-sequences) in the [Mg/Fe] vs. [Fe/H] plane are clearly visible for the APOGEE sample of Milky Way stars; the green dashed lines shown in the figure represent the median values of the low and high $\alpha$-sequences for metallicity bins of 0.09 dex. (The thin disk sample is defined as [Mg/Fe] $\leq$ 0.2 for [Fe/H] $\leq -$0.25 and [Mg/Fe] $\leq$ 0.15 for [Fe/H] $\geq -$0.25, while the thick disk stars as [Mg/Fe] $>$ 0.2 for [Fe/H] $\leq -$0.25 and [Mg/Fe] $>$ 0.15 for [Fe/H] $\leq -$0.25. 
We also show as the full orange line the linear least-squares fit for all the TriAnd stars (from this study and the literature), which indicates a slight increase in [Mg/Fe] with the decrease in metallicity ([Mg/Fe] = $-$0.082[Fe/H] $-$0.083). Such a pattern is reminiscent of an extension to lower metallicities of the radial [Mg/Fe] trend of more local thin disk stars (as proposed by \citealt{Hayes2018}) and this trend would pass through the TriAnd stars having disk orbits as indicated by their kinematics (\citealt{SalesSilva2019}).
It is noted, however, that the lowest metallicity TriAnd star from the \citet{Hayes2018} sample (2M00591640+3856025, with [Fe/H]= $-$1.11) has a [Mg/Fe] ratio of 0.08 dex, indicating an approximately constant [Mg/Fe] behavior of the TriAnd population in APOGEE DR14 (dashed orange line in the bottom panel of Figure \ref{singlealphaartigo5final}). 

Most interestingly, it can be seen that the results for the TriAnd stars analyzed in this study (shown as red circles), which are more metal-poor than those previously studied, indicate roughly a parallel sequence to the Milky Way local disk at lower values of [Fe/H]: the stars with the lowest metallicity in our TriAnd sample seem to indicate a more accentuated growth of the [Mg/Fe] ratio with the decrease of metallicity (dashed red line in the panel bottom Figure \ref{singlealphaartigo5final}) indicating that the TriAnd population presents a shifted 'knee' towards lower metallicities when compared to that characterized by the local Milky Way disk. This shifted 'knee' can be a consequence of the overall decrease in metallicity with an increase in Galactocentric distance and height above the plane, as predicted by models of chemical evolution of the Galaxy (e.g., \citealt{Minchev2014}, \citealt{Kubryk2015}). It is observed that the [$\alpha$/Fe] versus [Fe/H] pattern for Milky Way field stars in the APOGEE survey varies with Galactocentric distance and across the Galaxy (e.g., \citealt{Hayden2015}, \citealt{Queiroz2019}). 
Such behavior could be typical of the very outer disk, which has hardly been proved up to now; a comparison of the TriAnd results obtained here with the disk behavior at the outskirts of the Galaxy will only be possible with future APOGEE data releases and  the results of other Galactic surveys.



    \subsection{The [$\alpha_{h/ex}$] ratio}
    
The production of $\alpha$-elements in massive stars can happen either during the hydrostatic burning of C and Ne or during the SN II explosion (\citealt{WoosleyWeaver1995}); although much more reduced than in SN II, different $\alpha$ elements may have varying levels of production yields in SNe Ia (\citealt{Iwamoto1999}; \citealt{Maeda2010}). 
Possible differences between the abundance patterns of the $\alpha$ elements, for example, those formed by the hydrostatic (like Mg and O) and the explosive process (like Si, Ca and Ti) and, in particular, the hydrostatic/explosive (h/ex) $\alpha$ element ratio may provide valuable insights on the chemical evolution of a population (\citealt{Carlin2018}; \citealt{Blancato2019}).
   
\begin{figure*}
\centering
\includegraphics[width=15cm]{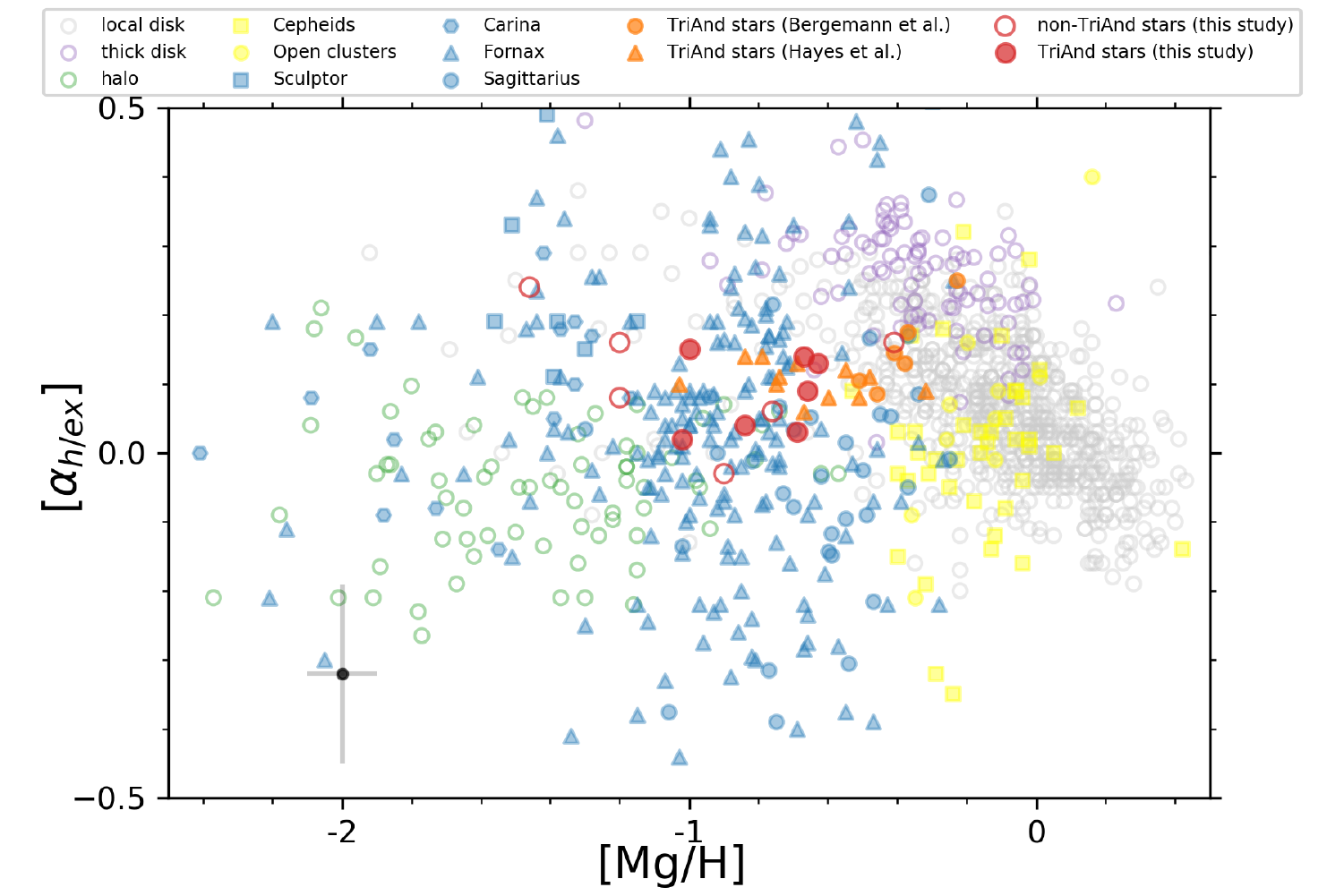}
\caption{Ratio between the $\alpha$-elements formed by hydrostatic (Mg and O) and the explosive process (Si, Ca and Ti) versus [Mg/H]. Symbols have the same meaning as in Figure \ref{singlealphaartigo4final}.}%
\label{singlealphaFieldartigofinal}
\end{figure*}

In Figure \ref{singlealphaFieldartigofinal} we show the [$\alpha_{h/ex}$] ratios versus [Mg/H] ratio, where $\alpha_{hydrostatic}$=([Mg/Fe] + [O/Fe])/2 and $\alpha_{explosive}$=([Si/Fe] + [Ca/Fe] + [Ti/Fe])/3. 
The samples are the same as before and correspond to stars in the TriAnd overdensity, along with the Milky Way disk, halo, and dwarf galaxies.
It is clear from Figure \ref{singlealphaFieldartigofinal} that the studied sample of TriAnd stars (red points) has a slightly positive $\alpha_{h/ex}$ ratio with a relatively small dispersion, having a mean [$\alpha_{h/ex}$]=0.08$\pm$0.05 (or, 0.02$\pm$0.06 if NLTE corrections were adopted for the derived LTE abundances of Ca and Ti, Table \ref{table:parameters}); a similar [$\alpha_{h/ex}$] ratio is also found in \citet{Hayes2018} ($\sim$0.1 $\pm$ 0.1 dex) and \citet{Bergemann2018}, except for one star in their sample that has a higher ratio. 
All in all, the TriAnd results are displaced in Mg from the local Milky Way disk (grey points) and thick disk (purple points). 
\citet{Carlin2018} discussed that the [$\alpha_{h/ex}$] ratios for dwarf galaxies were displaced from the Milky Way trend. In particular, they modelled this result for the Sagittarius galaxy and concluded that such [$\alpha_{h/ex}$] ratios were consistent with an IMF that lacked the most massive stars.
The TriAnd results fall in the parameter space that is also occupied by stars in dwarf galaxies (blue symbols). However, the dispersion in the [$\alpha_{h/ex}$] ratios for dwarf spheroidals is much larger probably due to the decay of this ratio in an approximately constant [Mg/H] ratio caused by the little enrichment in Mg after this decay starts (e.g., \citealt{Carlin2018}).

\section{Conclusions}

In \citet{SalesSilva2019} we conducted our first analysis of stars in the TriAnd overdensity; the studied sample covered a large range in metallicities, from $-$1.34$\leq$[Fe/H]$\leq-$0.78. We analyzed the stellar kinematics (orbits, proper motions and radial velocities) along with the chemical abundances of the elements Na, Al, Ni, Cr, Fe, Ba, and Eu. \citet{SalesSilva2019} concluded that the TriAnd population had a unique chemical pattern, not resembling any known population (dwarf galaxies, local disk, and halo). 
The results for the TriAnd stellar kinematics in that study indicated that its origin can be linked to the outer Galactic disk. 
In this study, we investigate the chemical pattern of five $\alpha$ elements in the same sample of TriAnd stars analyzed previously.
Overall, our results for the TriAnd overdensity in this study corroborate our previous findings: the chemical pattern of $\alpha$-elements in the TriAnd overdensity also differs from the local (thin and thick) disk. The results obtained for the TriAnd overdensity can be summarized as follows:

1) In general, the TriAnd stars analyzed have $\alpha$-element abundances (except for Si) that are displaced to lower values of [X/Fe] for a given [Fe/H] when compared  to  the  results  for  the  local  Galactic trend; in particular, these stars fall between the local disk and the dwarf galaxies in the [$\alpha$/Fe] vs. [Fe/H] plane (Figure \ref{singlealphaartigo4final} and the top panel of Figure \ref{singlealphaartigo5final}), although for Si the TriAnd stars show a pattern that is more similar to the local disk stars, a behavior that is also seen in some of the stars in dwarf galaxies.
The kinematical criteria used in \citet{SalesSilva2019} to classify our stars as TriAnd or non-TriAnd stars also chemically segregated them by the $\alpha$ elements, with the TriAnd stars showing lower [X/Fe] ratios than non-TriAnd stars in a similar metallicity range; this indicates a clear distinction between the abundance pattern of the control sample (non-TriAnd) stars and the TriAnd stars, all analyzed homogeneously.

2) Our TriAnd sample has stars with [Fe/H] less than $-$1.2 dex, allowing us to probe the evolution of the $\alpha$-elements in the low metallicity regime. 
For [Fe/H]$<-$1.2, the TriAnd stars studied here show an intriguing chemical pattern: for Mg, Si and Ca, the [X/Fe] ratios indicate an increase with a decrease in metallicity, while for the [O/Fe], and [Ti/Fe] ratios, we find a slightly decreasing behavior relative to the more metal-rich TriAnd stars. 
For Mg, in particular, a linear least-squares fit to the results for the TriAnd population in the [Fe/H] vs. [Mg/Fe] plane in comparison with the tendency found for APOGEE DR14 abundance pattern for more local disk hints that the high [Mg/Fe] ratios for the lower metallicities TriAnd stars may correspond roughly to a parallel sequence at lower values of [Fe/H] relative to the Milky Way local disk, revealing a shifted 'knee', for lower metallicities, for the TriAnd population relative to the more local disk. Thus, in addition to the TriAnd population representing an extension of the disk radial abundance gradient for the $\alpha$-elements, as shown by \citet{Hayes2018}, our results indicate that the TriAnd overdensity may also characterize a new knee for the outer disk.

3) In the [$\alpha_{h/ex}$] vs. [Mg/H] plane, the TriAnd stars are displaced in Mg from the local disk pattern, staying in the parameter space of some dwarf galaxy stars. However, TriAnd stars show a relatively small dispersion in the [$\alpha_{h/ex}$] ratio unlike the dwarf galaxy populations that present a much larger dispersion.

The TriAnd overdensity presents a peculiar chemical pattern for the $\alpha$ elements, differing from the known stellar populations of our Galaxy. Comparisons with the disk field population at R$_{GC} >$ 15 kpc is not yet possible because the composition of the remote disk is still not known. 
In addition, the small number of TriAnd stars with determined chemical abundances, particularly with [Fe/H]$<-$1, also limits our present ability to understand the puzzling chemistry of TriAnd and place it into an evolutionary and Galactic structure context.  Clearly more high resolution spectroscopic observations of additional TriAnd members at all metallicities is needed.

\acknowledgments

We thank Allyson Sheffield for providing the metallicity data used in \citet{Sheffield2014}. We thank Chris Hayes for extensive discussions. KC also thanks Kathryn Johnston and Verne Smith for discussions.
JVSS thanks FAPERJ proc. 202.756/2016. JVSS, HDP, HJR-P and FA-F thank the Brazilian Agency CAPES for the financial support of this research. HDP thanks FAPESP proc. 2018/21250-9.
This research has made use of the services of the ESO Science Archive Facility.

%

\vspace{5mm}
\facilities{Gemini North:GRACES, ESO-Archive:VLT:UVES}


\software{IRAF 
          Opera 
          }

\end{document}